\documentclass[12pt, nofootinbib]{revtex4-1}	
\usepackage{amsmath,amssymb,bbold}
\usepackage{caption}
\usepackage{tikz-feynman}
\usepackage{float}
\usepackage[makeroom]{cancel} 

\usepackage{xcolor}
\usepackage{ulem}

\newcommand{\be}{\begin{equation}}
\newcommand{\ee}{\end{equation}}
\newcommand{\bea}{\begin{eqnarray}}
\newcommand{\eea}{\end{eqnarray}}

\begin{document}
\title{Consequences of Minimal Entanglement in Bosonic Field Theories}
\author{Spencer Chang, Gabriel Jacobo}
\address{
Department of Physics and Institute for Fundamental Science\\ 
University of Oregon, Eugene, Oregon 97403}

\begin{abstract}
In this paper, we study a recently discovered connection between scattering that minimally entangles and emergent symmetries.  In a perturbative expansion, we have generalized the constraints of minimal entanglement scattering, beyond qubits, to general qudits of dimension $d.$  Interestingly, projecting on any qubit subspaces, the constraints factorize, so that it is consistent to analyze minimal entanglement by looking at all such subspaces.  We start by looking at toy models with two scalar fields, finding that minimal entanglement only allows quartic couplings which have instabilities at large field values and no symmetries.  For the two Higgs doublet model, by considering $H^+ H^-\to H^+ H^-$ scattering, we show that minimal entanglement in this channel does not allow an interacting parameter point with enhanced symmetries.  These results show that the connection between minimal entanglement and symmetries depends strongly on the scattering channels analyzed and we speculate on the potential resolutions.  

\end{abstract}

\maketitle

\section{Introduction}
 Symmetries play a crucial role in our understanding of particle spectra and interactions, so new techniques to understand them could be extremely useful.   Recently, analyses in specific contexts have shown that quantum entanglement constraints may lead to the emergence of symmetries.  For example, studying nonrelativistic baryon scattering has demonstrated that at parameter points where the scattering has minimal entanglement, there are enhanced symmetries \cite{Beane:2018oxh, Low:2021ufv, Liu:2022grf, Liu:2023bnr}. Similarly, this principle has been applied to interactions between relativistic scalar fields. In \cite{Carena:2023vjc}, it was found that a two Higgs doublet model (2HDM)  exhibits an enhanced $SO(8)$ global symmetry after demanding entanglement suppression for the scattering of $H^0_k H^+_l \to H^0_{i} H^+_{j}$, where the $i, j,k,l$ denote different flavors.  


By considering the S-matrix as a unitary transformation, one can impose entanglement suppression by averaging  entanglement measures over initial states.  Existing work has analyzed situations where the Hilbert space is effectively a two ``qubit" system, which in the baryon case is spin and in the 2HDM is flavor.  For such cases, analyzing $2\rightarrow 2$ scattering, it was demonstrated in \cite{Low:2021ufv} that there are only two minimally entangling logic gates allowed for the S-matrix, which are the Identity and the SWAP, where SWAP interchanges the states of the two qubits.  Since $S=\mathbb{1}+i T$, in theories with a perturbative coupling, the Identity S-matrix is the only allowed solution (up to change of basis of the qubits).

Given the potential new avenues that this symmetry and entanglement suppression connection may open, it is important to investigate this further. In this work, we focus on interactions between scalar fields in toy models and the two Higgs doublet model.  We will also briefly consider interactions between gauge bosons. The structure of the remainder of this paper is as follows: In Section \ref{sec:minimal_ent}, we describe the constraints on the $M$-matrix elements that our models must satisfy in order to achieve minimal entanglement.  These are derived for general qudits of dimension $d$, but the constraints for $d=2$ agree with the results in \cite{Carena:2023vjc}, showing that it is consistent to project onto two dimensional qubit subspaces. These constraints form the foundational basis for the subsequent analyses, where we will consider different choices for the qubits. Section \ref{sec:toyscalarmodels} explores the application of these constraints within simple toy models, involving real and complex scalar fields. Imposing minimal entanglement on these models, we find interactions that do not admit a compact symmetry, where each one has an instability at large field values. In Section \ref{sec:2HDM_gaugebosons}, we extend the analysis to more complex scenarios. Here, we inspect a 2HDM presented in \cite{Carena:2023vjc}, showing results from  different interaction channels that were not considered there.  By looking at different choices for the qubits in $2\to 2$ scattering, we find that entanglement suppression does not lead to interacting parameter points with enhanced symmetry, in agreement with the complementary analysis of \cite{Kowalska:2024kbs}.  
These results show that requiring entanglement suppression in all channels can be too constraining and lead to theories with either no interactions or instabilities.
Additionally in this section, we consider a non-abelian gauge theory to address concerns that $H^+H^-\to H^+H^-$ is related by crossing to $H^+ H^+ \to H^+ H^+$, which due to Bose symmetry may always have some entanglement power.  The gauge theory results suggest that this is not a concern.  In Section \ref{sec:conclusions}, we conclude and give some speculations on the potential consequences of our results.  Finally, in the Appendix, we give details on the scattering states we used and how to derive the minimal entanglement constraints.  
\section{General Minimal Entanglement Conditions}
\label{sec:minimal_ent}
In this paper, we will study the scattering of $|kl\rangle \to |i j\rangle$, where $i,j,k,l$ are discrete indices ranging $1, \cdots, d$.  These states are closely related to $s$-wave combinations of bosons, where the $i, j, k, l$ degrees of freedom are Kronecker delta normalized.  For those interested, the precise form of the states is given in the Appendix. We will study situations where the scattering minimally entangles two particles in the final state subsystems. This is equivalent to the constraint that any incoming product state remains a product state after the scattering. For an 
initial product state $|\psi_{initial}\rangle = a_{k} |k\rangle \otimes b_{l}|l\rangle$, with normalization $\sum_k |a_k|^2 = \sum_l |b_l|^2=1$, the final state is 
\begin{align}
|\psi_{final}\rangle=
S |\psi_{initial}\rangle = (\mathbb{1}+i T) a_{k} |k\rangle \otimes b_{l}|l\rangle =  a_k b_l(\delta_{ik}\delta_{jl}+i M_{ij,kl})|i\rangle \otimes |j\rangle = c_{i j} |i\rangle \otimes |j\rangle
\end{align}
so $c_{i j}= (\delta_{ik}\delta_{jl}+i M_{ij,kl})a_k b_l$.\footnote{Note that we order the indices for the initial and final states in the opposite way compared to  \cite{Carena:2023vjc}. However, all of our scattering amplitudes and the constraints of them are symmetric in exchanging initial and final state indices so this convention doesn't lead to any different consequences.} As shown in the Appendix, these amplitudes can be evaluated as integrals of the standard scattering amplitudes evaluated in the center of mass frame, 
\begin{align}
&  M_{ij,kl} = \nonumber \\ &  \frac{\sqrt{|\vec{p_i}|_{cm} |\vec{p_k}|_{cm}}}{8\pi \sqrt{P^2}}  \int_0^1 d\cos\theta_3 \int_0^1 d\cos\theta_1 \int_0^{2\pi} \frac{d\phi_1}{2\pi}\int_0^{2\pi} \frac{d\phi_3}{2\pi} \;  M(\Phi_k,p_1; \Phi_l, p_2 \to \Phi_i,p_3; \Phi_j,p_4).
\end{align}
Note that the angular integrals require particles $i, k$ to move forward, which allows distinguishability of the initial state particles, even if $i$ and $j$ (or $k$ and $l$) are identical.  However, this comes as the cost of making these integrals more challenging to compute.
In this paper, we will calculate the standard scattering amplitude, but it is important to keep in mind that the constraints from minimal entanglement only apply to the integrated $M_{ij,kl}$ amplitudes.  After scattering, the final state has a density matrix $\rho = | \psi_{final} \rangle \langle \psi_{final}|.$  Now let's consider the entanglement properties of the final state, by constructing the reduced density matrix $\rho_1=\text{Tr}_2 (\rho).$ Since the scattering is unitary, $\text{Tr}_1(\rho_1)=1$ and in its eigenstate basis, it has positive eigenvalues which sum to 1.  If scattering is minimally entangling, the final state is a product state, so $\rho_1$ has one nonzero eigenvalue of 1 and the rest being zero.  Thus, minimal entanglement requires in the eigenstate basis that the reduced density matrix $\rho_1$ satisfies $\rho_1^n  =\rho_1$.  One can write this in a basis independent way, via $\text{Tr}_1(\rho_1
^n) = \text{Tr}(\rho_1)=1=\text{Tr}_1(\rho_1)^n$ for $n=1, \cdots, d.$  This allows one to construct a set of entanglement measures
\begin{align}
X_n \equiv \text{Tr}(\rho_1)^n-\text{Tr}(\rho_1^n). 
\end{align}
$X_n$ have the property that they vanish if there is no entanglement for the final state and they have the maximum of $1-1/d^{n-1}$ when the final state is maximally entangled.  In fact, $X_2=0$ is closely related to the generalized concurrence proposed in \cite{PhysRevA.64.042315}.  
 
Imposing the constraint $X_2= 0$, we find that as a power series in the amplitudes $M$, $X_2$ starts at $O(M^2)$.  Since we will focus on theories that have a perturbative expansion, we will take the leading order result, giving the constraints 
\bea
&&M_{ij,kl} =0  \ \text{if} \ i \neq k, j \neq l, \label{eq:c1}\\
&&M_{ij,il} = M_{kj,kl} \ \text{if} \  j \neq l,\quad  M_{ij,lj} = M_{ik,lk}, \ \text{if} \  i \neq l \label{eq:c2}\\
&&M_{ij,ij} + M_{kl,kl} = M_{il,il} + M_{kj,kj}. \label{eq:c3}
\eea
One can also derive these constraints by requiring that the scattering has minimal entangling power when averaged over the initial state.  Details of these two calculation approaches to determine these constraints are given in the Appendix.  We've also observed that other $X_n=0$ constraints give the same constraints for various values of $d$, but do not claim to have proved this.  Note that these results are true for any number of states so that if we consider $i,j,k,l$ with just values 1 and 2, their amplitude constraints are independent of other possible scattering states. Restricting to these 2-dimensional subspaces, we recover the constraints on the $M$ matrix elements found in \cite{Carena:2023vjc}:
\bea
&&M_{11,22} = M_{12,21} = M_{21,12} = M_{22,11}=0,\label{condition2}\\
&&M_{11,12} = M_{21,22},\quad \ M_{11,21} = M_{12,22},\label{condition3}\\
&&M_{11,11} + M_{22,22} = M_{12,12} + M_{21,21}. \label{condition1}
\eea
The results in that paper assumed $d=2$ and utilized an entanglement measure called concurrence \cite{PhysRevLett.78.5022, PhysRevLett.80.2245}, which in our notation is $\Delta=|\text{Det}{\; c_{ij}}|.$  For $d>2$, concurrence is no longer a good entanglement measure, since it can vanish without the final state being a product state.  For example, with $d=3$, $\text{Det}\; {c_{ij}}$ vanishes if the matrix is rank 2, but the final state being a product state requires $c_{ij}$ is rank 1.  However, imposing the $X_n=0$ constraints work for any $d$ and thus are more generally applicable than concurrence.  

We will mostly work with two-state subsystems and will use the matrix notation
\begin{align}
    M=\begin{pmatrix}
      M(11\to 11) & M(11\to 12) & M(11\to 21) & M(11\to 22)\\
      M(12\to 11) & M(12\to 12) & M(12\to 21) & M(12\to 22)\\
      M(21\to 11) & M(21\to 12) & M(21\to 21) & M(21\to 22)\\
      M(22\to 11) & M(22\to 12) & M(22\to 21) & M(22\to 22)
    \end{pmatrix}
\end{align}
to summarize our calculations of the standard amplitudes.  As discussed above, the minimal entanglement constraints apply to these matrix elements integrated over the $t$ Mandelstam variable.  When the $M(kl\to ij)$ matrix elements depend only on $s$, one does not have to integrate, since the $t$ integral cannot vanish.  We will discuss cases which have $t$ dependence more carefully.  For most of our scenarios, Eqs.~(\ref{condition2}) and (\ref{condition1})  will give nontrivial constraints, while Eqs.~(\ref{condition3}) are automatically satisfied.  To ease interpretation, we note that Eq.~(\ref{condition2}) shows that the anti-diagonal of the scattering matrix vanishes after integration, while Eq.~(\ref{condition1}) says that the integrals of the sum of the first and last diagonal elements equals the integral of the sum of the other two diagonal elements.  Thus, we will describe the constraint of Eq.~(\ref{condition1}) as the on-diagonal constraint.      

In some of the analysis that follows, we will show that by looking at all scattering channels, it may be impossible to satisfy the minimal entanglement constraints for nonzero $M_{ij,kl}.$  However, since all entanglement measures vanish for the case of a free theory, the free theory is always an allowed solution for minimal entanglement.  In these cases, these show that minimal entanglement forces a perturbative theory to become free, which has enhanced symmetries but at the cost of being trivial. 

\section{Minimal Entanglement for Toy Models with two  scalars}
\label{sec:toyscalarmodels}
Consider the scattering of two particles where our spectrum contains two real scalars $\phi_1, \phi_2$.  
%
%
Let's consider the quartic couplings ${\cal L} \supset -\lambda_{11} \phi_1^4 -\lambda_{12} \phi_1^2 \phi_2^2 - \lambda_{22} \phi_2^4.$  For $i,j,k,l=(\phi_1,\phi_2)$, one finds the scattering amplitudes:
\begin{align}
    M=-\begin{pmatrix}
      24\lambda_{11} & 0 & 0 & 4\lambda_{12}\\
      0 & 4\lambda_{12} & 4\lambda_{12} & 0\\
      0 & 4\lambda_{12} & 4\lambda_{12} & 0\\
      4\lambda_{12} & 0 & 0 & 24\lambda_{22}
    \end{pmatrix}.
\end{align}
Imposing the entanglement suppression conditions, the anti-diagonal vanishing requires $\lambda_{12}=0$.  Then imposing that condition on the diagonal requires $24\lambda_{11}+24\lambda_{22}=4\lambda_{12}+4\lambda_{12}=0$ or $\lambda_{22}=-\lambda_{11}$.   If we had instead allowed all possible quartics, we would have found that entanglement suppression allows the potential 
\begin{align}
V= \lambda (\phi_1^4-\phi_2^4) + \kappa (\phi_1^2+\phi_2^2)\phi_1 \phi_2.
\end{align}
These quartic couplings admit no compact symmetry for the real scalars. The $\lambda$ coupling is symmetric under a hyperbolic rotation  $(\phi_1^2,\phi_2^2) \to \sqrt{\phi_1^4-\phi_2^4}\; (\cosh \eta, \sinh \eta)$, but both types of couplings have no stable global minimum with instabilities at large field values.  Also, the hyperbolic transformation is not a symmetry of the kinetic term.\footnote{It is interesting to note that such a hyperbolic symmetry of the potential  has been used to create models with natural electroweak symmetry breaking, e.g.~\cite{Cohen:2018mgv}.} 
In this case, minimal entanglement coupled with the requirement of a stable vacuum requires all quartic interactions to vanish.  
%

Now let's consider the next simplest case, where  $\phi_1$, $\phi_2$ are complex scalars, with separate $U(1)$ symmetries.  We consider the general quartic interactions ${\cal L} \supset -\lambda_{11} |\phi_1|^4 -\lambda_{12} |\phi_1|^2 |\phi_2|^2 - \lambda_{22} |\phi_2|^4$.  Unlike the real scalar case, we can analyze different scattering processes by choosing different qubit subsystems for $i, j, k, l$ as shown in Table~\ref{tbl:ijchoices}.
\begin{table}[h!]
\begin{center}
\begin{tabular}{|c|c|c|}
\hline
& $i, k$ & $j, l$\\ \hline
$1)$ & $\phi_1$, $\phi_1^*$ & $\phi_2$, $\phi_2^*$\\
$2)$ & $\phi_1$, $\phi_2$ & $\phi_1^*$, $\phi_2^*$\\
$3)$ & $\phi_1$, $\phi_2^*$ & $\phi_1^*$, $\phi_2$\\
\hline
\end{tabular}
\caption{
Allowed choices for the two ``qubit" degrees of freedom in $i, j, k, l$. \label{tbl:ijchoices}}
\end{center}
\end{table}
For example, for case $1)$, the allowed initial and final states are $| \phi_1 \phi_2 \rangle$, $| \phi_1 \phi_2^* \rangle$, $| \phi_1^* \phi_2 \rangle$ and $| \phi_1^* \phi_2^* \rangle$. The corresponding $M$-matrices for the three cases are: 
\newline
\noindent Case $1)$: $i, k=(\phi_1, \phi_1^*)$,  $j,l=(\phi_2, \phi_2^*)$\\
\begin{align}
M=-\begin{pmatrix}
    \lambda_{12} & 0 & 0 & 0 \\
    0 & \lambda_{12} & 0 & 0 \\
    0 & 0 & \lambda_{12} & 0 \\
    0 & 0 & 0 & \lambda_{12}
\end{pmatrix},
\end{align}
\noindent Case $2)$: $i, k=(\phi_1, \phi_2)$, $j, l=(\phi_1^*, \phi_2^*)$ \\
\begin{align}
M=-\begin{pmatrix}
    4\lambda_{11} & 0 & 0 & \lambda_{12} \\
    0 & \lambda_{12} & 0 & 0 \\
    0 & 0 & \lambda_{12} & 0 \\
    \lambda_{12} & 0 & 0 & 4\lambda_{22}
\end{pmatrix},
\end{align}
\noindent Case $3)$: $i, k=(\phi_1, \phi_2^*)$, $j, l=(\phi_1^*, \phi_2)$ 
\begin{align}
M=-\begin{pmatrix}
    4\lambda_{11} & 0 & 0 & \lambda_{12} \\
    0 & \lambda_{12} & 0 & 0 \\
    0 & 0 & \lambda_{12} & 0 \\
    \lambda_{12} & 0 & 0 & 4\lambda_{22}
\end{pmatrix}.
\end{align}
We now impose the constraints on the scattering amplitudes to suppress entanglement. In case $1)$, all constraints are satisfied.  However, in cases $2)$ and $3)$, entanglement suppression is achieved if $\lambda_{12} = 0$ and $\lambda_{11}=-\lambda_{22}$, leading us to the interaction potential $\lambda (|\phi_1|^4 - |\phi_2|^4)$.  This potential is very similar to the first term of the real scalar case and again shows that by looking at the general allowed scattering states, we find a potential with no stable vacuum, which has a hyperbolic symmetry that is not consistent with the kinetic term.  Thus, there appear to be no simple two scalar particle field theories which have enhanced symmetries when imposing minimal entanglement, unless the theories are free.  

\section{Minimal Entanglement for Two Higgs doublet models and Non-abelian gauge bosons}
\label{sec:2HDM_gaugebosons}
In \cite{Carena:2023vjc}, the general CP preserving two Higgs doublet model was analyzed, for the scattering channel $H^+_k H^0_l \rightarrow H^+_{i} H^0_{j}$.\footnote{Reproducing the analysis in \cite{Carena:2023vjc}, we find the interpretation of $H^0_l$ is that one needs to consider all of the cases where $H^0_j$ is two of the neutral Higgs bosons $h_1^0, h_2^0, A_1^0, A_2^0$.} Imposing the entanglement suppression conditions for this channel, one finds the potential $V = \lambda \left(|H_1|^2 + |H_2|^2 -\dfrac{v^2}{2}\right)^2$, which has an enhanced $SO(8)$ symmetry.  This is of phenomenological interest, since softly breaking the $SO(8)$ symmetry leads to an approximate alignment limit for the lightest Higgs boson \cite{BhupalDev:2014bir, BhupalDev:2017txh}.  

However, as we will now show, this is because not all of the scattering channels were analyzed.  Of the fields in
\bea
H_1=\begin{pmatrix}
    H_1^+\\
    \dfrac{h_1+v+ i A^0_1}{\sqrt{2}}
\end{pmatrix}, \quad H_2=\begin{pmatrix}
    H_2^+\\
    \dfrac{h_2+ i A^0_2}{\sqrt{2}}
\end{pmatrix},
\eea
we consider the processes $H^+_k H^-_l \rightarrow H^+_{i} H^-_{j}$. The $M$ matrix for the 4-point interactions is 
\begin{align}
M_{\text{4-point}}=-\begin{pmatrix}
    4\lambda & 0 & 0 & 2\lambda \\
    0 & 2\lambda & 0 & 0 \\
    0 & 0 & 2\lambda & 0 \\
    2\lambda & 0 & 0 & 4\lambda
\end{pmatrix}.
\end{align}
Additionally, we have contributions from the $s$ and $t$-channels mediated by $h_1$:
\begin{align}
M_{s}=-\begin{pmatrix}
    \dfrac{4 \lambda^2 v^2}{s-2 \lambda v^2 } & 0 & 0 & \dfrac{4 \lambda^2 v^2}{s-2 \lambda v^2 } \\
    0 & 0 & 0 & 0 \\
    0 & 0 & 0 & 0 \\
    \dfrac{4 \lambda^2 v^2}{s-2 \lambda v^2 } & 0 & 0 & \dfrac{4 \lambda^2 v^2}{s-2 \lambda v^2 }
\end{pmatrix}, \ 
M_{t}=-\begin{pmatrix}
    \dfrac{4 \lambda^2 v^2}{t-2 \lambda v^2 } & 0 & 0 & 0 \\
    0 & \dfrac{4 \lambda^2 v^2}{t-2 \lambda v^2 } & 0 & 0 \\
    0 & 0 & \dfrac{4 \lambda^2 v^2}{t-2 \lambda v^2 } & 0 \\
    0 & 0 & 0 & \dfrac{4 \lambda^2 v^2}{t-2 \lambda v^2 }
\end{pmatrix}.
\end{align}
All together, then one gets
\begin{align}
M=-2\lambda  \begin{pmatrix}
 \dfrac{s}{s-2 \lambda v^2 }+\dfrac{t}{t-2 \lambda v^2 } & 0 & 0 &  \dfrac{s}{s-2 \lambda v^2 } \\
    0 & \dfrac{t}{t-2 \lambda v^2 } & 0 & 0 \\
    0 & 0 &\dfrac{t}{t-2 \lambda v^2 } & 0 \\
     \dfrac{s}{s-2 \lambda v^2 } & 0 & 0 &  \dfrac{s}{s-2 \lambda v^2 }+\dfrac{t}{t-2 \lambda v^2 }
\end{pmatrix}.
\end{align}
This form of the scattering matrix is expected since $H_1^+, H_2^+$ are both Nambu-Goldstone bosons for this potential, so that the amplitudes vanish in the limit $s,t\to 0$. Now, imposing  the integrated anti-diagonal condition, we either need $s=0$ or $\lambda =0$.  If one imposes $s=0$, then the on-diagonal condition is automatically satisfied.  Unfortunately, this low energy limit does not have a physical interpretation.  This is because $H_1^+$ is an exact eaten Nambu-Goldstone boson, so it is not a physical state; thus, we can at best use it, via the equivalence theorem, to approximate the scattering of longitudinal $W^+$ at $s,|t|\gg m_W^2$.   As for $H_2^+$, experimental constraints require it to have a mass above $\sim 100$ GeV  \cite{BhupalDev:2017txh}.   Since both of these considerations require $s \gtrsim (100 \text{ GeV})^2$,  the only minimal entanglement solution with a physical interpretation is $\lambda=0$, again requiring the theory to be free.

In addition to the interactions in the potential, the Higgs doublets also have gauge interactions which add other contributions.  For $H^+ H^-$ scattering, this adds  terms from $\gamma$ and $Z$ bosons propagating in the $s$ and $t$ channel.  These give corrections of the form 
\begin{align}
\delta M(kl\to ij)=e^2 \left[\delta_{ij}\delta_{kl}(u-t)\left(\dfrac{1}{s}+\dfrac{\cot(2\theta_W)^2}{s-m_Z^2}\right)+\delta_{ik}\delta_{jl}(u-s)\left(\dfrac{1}{t}+\dfrac{\cot(2\theta_W)^2}{t-m_Z^2}\right)\right].
\end{align}
Focusing on $H^+_2 H^-_2 \to H^+_1 H^-_1$, we get a correction $\delta M(22\to 11) = e^2 (u-t)\left(\dfrac{1}{s}+\dfrac{\cot(2\theta_W)^2}{s-m_Z^2}\right)$.  
After integrating the total matrix element, we find
\begin{align}
M_{11,22}= -\frac{1}{16\pi} \left[\frac{e^2}{4} \left(1+\frac{\cot(2\theta_W)^2 s}{s-m_Z^2}\right)+\frac{2 \lambda s}{s-2\lambda v^2}\right].
\end{align}
This vanishes at two center of masses, $s\sim 35^2, 92^2\; \text{GeV}^2 \lesssim m_W^2$, none of which are large enough to have a physical interpretation.
  Thus, taking into account the gauge interactions does not fix the issues above.  Finally, we have also considered the general two Higgs doublet model potential and worked out the entanglement suppression conditions from $H^+_k H^-_l \to H^+_{i} H^-_{j}$ alone.  Even by restricting to this channel alone, we still find that all of the quartic couplings in the potential must vanish.  Note that similar conclusions to this was found using momentum wave packet scattering in \cite{Kowalska:2024kbs}.
%

One criticism of the $H^+ H^- \to H^+ H^-$ scattering channel is that by crossing symmetry, this is related to $H^+ H^+ \to H^+ H^+$ scattering, which due to Bose symmetry will tend to lead to some amount of entanglement.\footnote{We thank I.~Low for discussions on this point.}  To investigate this, we will now consider an analogous scenario involving gauge boson scattering.  
Considering the Lagrangian of a general non-abelian gauge theory, ${\cal L} \supset -\frac{1}{4} (\partial_\mu A_\nu^a -\partial_\nu A_\mu^a +g f^{abc} A_\mu^b A_\nu^c)^2$  where $f^{abc}$ are the structure constants and $g$ the gauge coupling, we studied the scattering process $A_k^a A_l^b \to A_{i}^a A_{j}^b$ where $i,j,k,l=$ L,R are the left and right-handed polarizations of the gauge bosons. The result at tree-level is
\begin{align}
M=2 g^2 f^{abc}f^{abc}  
\renewcommand\arraystretch{1.4}
\begin{pmatrix}
 \dfrac{s}{u} & 0 \ & 0 \ &  0 \\
    0 & \dfrac{u}{s} \ & \dfrac{t^2}{s u} \ & 0 \\
    0 & \dfrac{t^2}{s u} &\dfrac{u}{s} & 0 \\
    0 & 0 & 0 &  \dfrac{s}{u}
\end{pmatrix} \label{Gauge_Scattering}
\end{align}
with no sum on the $a,b$ index.  
For the anti-diagonal integral, the integrand has a fixed sign but the integral has a potential divergence when $u$ vanishes.  However, $u$ vanishes only if $\theta_i=\theta_k=\pi/2$ and $\phi_i-\phi_k=\pi$, which is restrictive enough for the integrals to converge (on the other hand, integrals of $1/t$ would have diverged).  Thus, from the anti-diagonal integral condition, minimal entanglement requires the gauge coupling to vanish, $g=0$.  An analysis of the diagonal condition results in the same requirement.
 Note that this scattering channel has the same criticism as the $H^+H^-$ channel, since it is related to $A^a A^a \to A^b A^b$ through crossing.  

Let's compare this to $W_k^+ Z_l \to W_{i}^+Z_{j}$ with $W^{\pm}=\frac{1}{\sqrt{2}}(A^1 \mp i A^2)$, $Z=A^3$ and $i,j,k,l=$ L,R for an unbroken $SU(2)$ gauge theory. Here, we get the same matrix (\ref{Gauge_Scattering}) where the $f^{abc}f^{abc}$ factor goes to 1.  $W^+ Z \to W^+ Z$ is the gauge boson analogue of $H^+ H^0 \to H^+ H^0$, without the crossing symmetry criticism, but we see that we get the same constraints as $A^a A^b\to A^a A^b$.  We take this as evidence that the crossing symmetry criticism is not generally valid.  
\section{Conclusions}
\label{sec:conclusions}
In this paper, we have generalized the constraints of minimal entanglement of \cite{Carena:2023vjc} beyond qubits to qudits of arbitrary dimension $d.$  One of the key results is that the generalized constraints factorize in a way that one can apply the constraints of \cite{Carena:2023vjc} to any two-dimensional subsector.

Using this result, we have shown in a few simple scalar field theories, that requiring entanglement suppression in all $2\to2$ channels does not lead to interacting theories with enhanced symmetries.  In a theory with either two real scalars or two complex scalars, the resulting quartic potential is unbounded from below.  The potential could have hyperbolic symmetry, but that is inconsistent with the kinetic term. Thus, the only allowed solution with enhanced symmetry is where the interactions vanish.  Extending the two Higgs doublet analysis of \cite{Carena:2023vjc} by analyzing $H^+_k H^-_l \rightarrow H^+_{i} H^-_{j}$ scattering, one finds that entanglement suppression in this channel requires the scalar potential to vanish, which as a free theory has an enhanced symmetry but again trivial dynamics.  These results show that there is more work needed to develop the connection between entanglement suppression and emergence of nontrivial enhanced symmetries.  We now speculate on a few possible resolutions of the raised issues.

First, due to the dependence on the scattering channels analyzed, an additional criteria may be needed to determine which scattering channels can be analyzed for minimal entanglement. Since a free theory is minimally entangling, it is always a candidate solution, so imposing the minimal entanglement conditions on all channels may restrict oneself to free theories.  As discussed above, channels where there are identical particles in the initial or final state (and those related by crossing) tend to entangle, so it might be correct to exclude these.  However, since the $A^a A^b \to A^a A^b$ constraints agreed with $W^+ Z \to W^+ Z$ and only the former has the crossing symmetry criticism, this gives an example where this criticism is incorrect.  Studying further examples may give some insights into the correct selection criteria.

Second, entanglement suppression may only lead to symmetric interacting theories for certain kinematics.  For instance, the nuclear theories with emergent symmetries had scattering that was non-relativistic \cite{Beane:2018oxh, Low:2021ufv, Liu:2022grf, Liu:2023bnr}.  In this work, we saw that the two Higgs doublet and the gauge theories satisfied minimal entanglement at threshold when $s\to 0.$  Again, more studied examples where symmetries emerge for only special kinematics would be useful.  

Finally, the connection may only hold in theories with certain other properties.  For instance, for strongly interacting theories (like nucleon scattering), our perturbative analysis would not apply.  Or perhaps it depends strongly on the particle content, such as only for fermions.  
%
To conclude, there remains much to understand in this interesting but still mysterious connection between minimal entanglement and emergent symmetries.  We hope to explore these and other directions in future work.  

\section*{Acknowledgements}
\label{sec:Acknowledgements}
The work of S.C.~and G.J.~was supported in part by the U.S. Department of Energy under Grant Number DE-SC0011640.  We thank P.~Asadi, G.~Kribs, and especially I.~Low for discussions.

\section*{Appendix}
\label{sec:appendix}
In this appendix, we will treat two issues of our calculation with care.  The first is the use of a basis for scattering which is very close to  $s$-wave states.  These are unit normalized states that allow for the definition of a standard unitary $S$-matrix.  Given this unitary matrix, we can then analyze the constraints from minimal entanglement.  We will do this in two ways, by looking at a generalization of concurrence and also the entanglement power of the unitary $S$-matrix.

\subsection{Scattering State Basis}
It is important to utilize unit-normalized states when discussing the entanglement properties of the $S$-matrix.  This is because this enables a normal unitary $S$-matrix to be defined for these states, which allows proper statistical interpretation of the entropy and other measures of entanglement.  

In this paper, we choose to define states that are closely related to $s$-wave states.\footnote{Another approach is to use momentum wavepackets, as studied recently \cite{Kowalska:2024kbs}.  However, the wavepacket approach has an undetermined parameter $\Delta$ which, as a ratio of delta functions with zero argument, is ill-defined.  }  Similar to \cite{Chang:2019vez}, we define states $|\alpha, P\rangle$ that have a discrete index $\alpha$ with total four momentum $P$, normalized so that $\langle \alpha, P | \beta, P'\rangle = \delta_{\alpha\beta}(2\pi)^4 \delta^4(P-P').$  For these states, $\langle \alpha, P | S |\beta, P'\rangle = (2\pi)^4 \delta^4(P-P') S_{\alpha\beta}$  where $S_{\alpha\beta}=\delta_{\alpha\beta}+i M_{\alpha\beta}$ is a normal unitary matrix and where the completeness relation is $\mathbb{1} = \sum_\alpha \int \frac{d^4P}{(2\pi)^4} |\alpha, P\rangle \langle \alpha, P|.$  

For this paper, we will define a 2-body subset of these states.  In the center of mass frame of $P$, we define
\begin{align}
|k, l, P\rangle = C_{kl} \int \frac{d^3 p_1}{(2\pi)^3 2 E_1}\frac{d^3 p_2}{(2\pi)^3 2 E_2} (2\pi)^4 \delta^4(P-p_1-p_2) \Theta(p_1^z)|\Phi_k, p_1; \Phi_l, p_2\rangle,
\end{align}
which consist of scalars of type $k, l$ with respective individual momentum $p_1, p_2$ and total momentum $P=p_1+p_2.$  Without the theta function, these are $s$-wave states containing two scalars.  The theta function ensures that the $k$ particle moves to the right (and from momentum conservation, the $l$ particle will move to the left).  Since moving left or right distinguishes those particles, this distinguishes the state $|k,l,P\rangle$ from $|l,k,P\rangle$ and will ensure that these give a product vector space of size $d^2$ if  $k,l =1, \cdots d.$  This mocks up a simple EPR-like setup where there are ideal detectors to the left and right, which  can perfectly measure the particles that come through them.  Note that if we were to consider scattering where $i, j$ cannot be the same particle, we could just integrate over the full angular range, resulting in the standard partial waves.

We can fix the normalization $C_{kl}$ by requiring $\langle i,j, P' | k,l, P\rangle = \delta_{ik}\delta_{jl}(2\pi)^4 \delta^4(P-P')$. This is satisfied if $C_{kl}= \sqrt{\frac{8\pi \sqrt{P^2}}{|\vec{p}_k|_{cm}}}$, which in the center of mass frame gives
\begin{align}
|k, l, P\rangle = \sqrt{\frac{|\vec{p_k}|_{cm}}{8\pi \sqrt{P^2}}} \int_0^1 d\cos\theta_1 \int_0^{2\pi} \frac{d\phi_1}{2\pi}\; |\Phi_k, p_1; \Phi_l, p_2\rangle.
\end{align}
From these states, we get matrix elements $\langle i, j, P' | S| k,l, P\rangle = S_{ij,kl}(2\pi)^4 \delta^4(P-P'),$ where $S_{ij,kl}=\delta_{ik}\delta_{jl}+i M_{ij,kl}.$  These scattering matrix elements can be determined from the usual matrix elements $M(kl \to ij)\equiv M(\Phi_k,p_1; \Phi_l,p_2; \to \Phi_i,p_3;\Phi_j,p_4)$ through
\begin{align}
M_{ij,kl} = \frac{\sqrt{|\vec{p_i}|_{cm} |\vec{p_k}|_{cm}}}{8\pi \sqrt{P^2}}  \int_0^1 d\cos\theta_3 \int_0^1 d\cos\theta_1 \int_0^{2\pi} \frac{d\phi_1}{2\pi}\int_0^{2\pi} \frac{d\phi_3}{2\pi} \; M(kl\to ij).
\end{align}
This integral can be performed by writing the matrix element in terms of the scattering angle between the particles $k$ and $i$, 
\begin{align}
\cos{\theta_{ik}} = \cos{\theta_i}\cos{\theta_k}+\cos(\phi_i-\phi_k)\sin{\theta_i}\sin{\theta_k}.
\end{align}
When there are massless particles being exchanged, these integrated amplitudes have potential divergences.  As discussed in the main text, integrals of $1/u$ are convergent while those of $1/t$ are divergent.  In the latter case, it may be possible to regularize the integrals, using ideas developed for Coulomb scattering, e.g.~\cite{Taylor:1973da}.

\subsection{Generalized Concurrence and Entanglement Power}
Consider the initial state $|\psi_{initial}\rangle = a_{k} |k\rangle \otimes b_{l}|l\rangle$ where  $\sum_k |a_k|^2 =\sum_l |b_l|^2=1$. The final state is $|\psi_{final}\rangle= S |\psi_{initial}\rangle = c_{i j} |i\rangle \otimes |j\rangle$ where  $S=\mathbb{1}+i T$. For minimal entanglement, we want the density matrix to be separable,
\be
\rho=| \psi_{final} \rangle \langle \psi_{final}|\quad  {\overset {\text{if min.ent.}} {=}} \quad | \phi_1 \phi_2 \rangle \langle \phi_1 \phi_2|= | \phi_1 \rangle \langle \phi_1 | \otimes | \phi_2 \rangle \langle \phi_2 | = \rho_1 \otimes \rho_2 \label{eq:rho1_min}
\ee
The reduced density matrix $\rho_1$ is obtained from $\rho_1 = \text{Tr}_2 [\rho]$, or
\begin{align}
\rho_1 = \sum_{j = 1}^d c_{ij} c_{kj}^* |i\rangle\langle k| \quad \text{where } c_{i j}= (\delta_{ik}\delta_{jl}+i M_{ij,kl})a_k b_l. \label{eq:rho1exp}
\end{align}
As show in Eq.~\ref{eq:rho1_min}, if $S$ is minimally entangling, $\rho_1 =|\phi_1\rangle \langle \phi_1| $ is a rank 1 matrix.  Since $\rho_1$ has positive eigenvalues which sum to 1, a basis-independent criterion for this is that $1^n=\text{Tr}_1 [\rho_1]^n = \text{Tr}_1 [\rho_1^n]$ for $n=2, \cdots, d$ (and also $\text{Det}(\rho_1)=0$). So we can define  
\be
X_n \equiv \text{Tr}_1 [\rho_1]^n - \text{Tr}_1 [\rho_1^n].
\label{TraceEquation}
\ee
These quantities are the natural generalizations of concurrence $\Delta=|\text{Det}\; c_{ij}|$, which is only an entanglement measure for $d=2.$  First of all, $\text{Tr}_1(\rho_1^n) = 1$ if $|\Psi_{final}\rangle$ is a product state, so $X_n=0$ if there is no entanglement.  $X_n$ are also maximized when $\rho_1$ is a diagonal matrix with entries of $1/d$, which represents maximal entanglement, so we have   $0\leq X_n \leq 1- 1/d^{n-1}$  where $X_n$  monotonically tracks the entanglement.  In particular, $X_2$ is proportional to the generalized concurrence in \cite{PhysRevA.64.042315} and will be the one we typically study.  

Let's use the expression above with $n=2$ when $d=2$, i.e.~a two-state system.\footnote{We have checked for this and some other $d$, that we get identical constraints for other values of $n$.} We can track the terms with a certain order in $M$-matrix elements by considering $S=\mathbb{1}+i \lambda T$, where $\lambda$ is just a real parameter whose power will indicate the order, e.g terms proportional to $\lambda^m$ are of order $O(M^m)$. The result is
\begin{align}
0 & = \lambda^2 \left|  a_1 a_2 b_1 b_2 (M_{12,12} + M_{21,21} - M_{11,11} - M_{22,22}) \right. \nonumber\\
& \quad +  a_1 a_2 b_1^2 (M_{11,12} - M_{21,22})
+ a_1 a_2 b_2^2 (M_{22,21} - M_{12,11})  \\
& \quad + a_1^2 b_1 b_2 (M_{11,21} - M_{12,22})+ a_2^2 b_1 b_2 (M_{22,12} - M_{21,11}) \nonumber \\
& \quad \left. + a_2^2 b_1^2 M_{21,12} - a_2^2 b_2^2 M_{22,11} - a_1^2 b_1^2 M_{11,22} + a_1^2 b_2^2 M_{12,21} \right|^2 + \mathcal{O}(\lambda^3). \nonumber
\end{align} 
Working with leading order terms and assuming that entanglement vanishes for any initial configuration, each factor inside the parenthesis and the $M$-matrix elements in the last line must vanish, leading to
\bea
&& M_{11,22} = M_{12,21} = M_{21,12} = 0, \label{type1} \\
&& M_{11,12} = M_{21,22}, \ M_{11,21} = M_{12,22}, \ M_{22,12} = M_{21,11}, \ M_{22,21} = M_{12,11}, \label{type2}\\
&&M_{11,11} + M_{22,22} = M_{12,12} + M_{21,21}. \label{type3}
\eea
Note that we have three classes of constraints: Type-1 constraints are characterized by the first equation and can be simplified as $M_{ij,kl} =0  \ \ \text{where} \ i \neq k$ and $ j \neq l$. Type-2 constraints are characterized by the second equation and can be simplified as $M_{ij,il} = M_{kj,kl} \ \text{with} \  j \neq l$, $ M_{ij,lj} = M_{ik,lk} \ \text{with} \  i \neq l$. Finally, Type-3 constraint is characterized by the third equation and can be written as $M_{ij,ij} + M_{kl,kl} = M_{il,il} + M_{kj,kj}.$
One might wonder whether the existence of additional states would affect the relations (\ref{type1}-\ref{type3}), but we can verify that this is not the case. Given $d=3$, we proceed with an analysis analogous to the previous scenario. Since (\ref{TraceEquation}) now yields many more terms, it might be convenient to use
\be
\dfrac{\partial^m}{\partial \lambda^m} \left. \left(\text{Tr}_1 [\rho_1]^2 - \text{Tr}_1 [\rho_1^2] \right) \right |_{\lambda=0}=0
\label{TraceEquation2}
\ee
This equation is automatically satisfied for $m=0,1$. The leading order terms appear when $m=2$, and the result looks like

\be
0 = |a_1|^4 |b_1|^4 (|M_{11,22}|^2 +|M_{11,23}|^2 + |M_{11,32}|^2 + |M_{11,33}|^2)+...
\ee
where the rest of the terms are also factorized in $a_i,a_i^*,b_j,b_j^*$. We know that the factor in the parenthesis must be zero. Since the sum of squared real terms is zero only if each term is zero, we can conclude that $M_{11,22}=M_{11,23}=M_{11,32}=M_{11,33}=0$. When $i,j=1,2$, the term proportional to $|a_1|^4 |b_1|^4$ was just $|M_{11,22}|^2$. Therefore, we can see that adding more states only introduces additional constraints for $M_{11,23}, M_{11,32},M_{11,33}$ without altering the one for $M_{11,22}$. Additionally, note that $M_{11,23}$,  $M_{11,32}$ and $M_{11,33}$ vanishing also belong to the Type-1 class mentioned previously. In fact, by doing the full analysis, we find that
\bea
&&M_{ij,kl} =0  \ \text{if} \ i \neq k, j \neq l,\\
&&M_{ij,il} = M_{kj,kl} \ \text{if} \  j \neq l, \quad M_{ij,lj} = M_{ik,lk} \ \text{if} \  i \neq l\\
&&M_{ij,ij} + M_{kl,kl} = M_{il,il} + M_{kj,kj},
\eea
where the indices run from 1 to 3.  One can see from this form that adding degrees of freedom only adds extra constraints, but does not modify the ones already found for $d=2.$


We can also understand these constraints from the point of view of the $S$ matrix having zero entangling power.  The entanglement power can be defined by averaging the initial product state $|\Psi_{initial}\rangle = a_k |k\rangle \times b_l |l\rangle$, where \cite{Beane:2018oxh}
\begin{align}    
{\cal{E}}(S) = 1- \int d{\bf a}\; d{\bf b}\; \text{Tr}_1(\rho_1^2) =  \int d{\bf a}\; d{\bf b}\; \left[\text{Tr}_1(\rho_1)^2-\text{Tr}_1(\rho_1^2)\right]
\end{align}
and $\int d{\bf a}, d{\bf b}$ are the integration measures when averaging over the initial states $a_k |k\rangle$, $b_l |l\rangle.$  Note that in the final equality, we've used the fact that scattering is unitary, so $\text{Tr}_1(\rho_1)=1.$  This choice is convenient because as we've seen above,  
$\text{Tr}_1(\rho_1)^2-\text{Tr}_1(\rho_1^2)$ starts at quadratic order in the matrix elements $M.$  To evaluate the integrals, we utilize $SU(d)$ symmetry to restrict the integrals to the form
\begin{align}
\int d{\bf a}\;  a_i a_j^* = \frac{1}{d} \delta_{ij},\quad \int d{\bf a}\;  a_i a_j a_k^* a_l^* = \frac{1}{d(d+1)} (\delta_{ik}\delta_{jl}+ \delta_{il}\delta_{jk}),
\end{align}
and the overall coefficients are determined by the fact that the states are unit normalized, $\int d{\bf a} \; a_i a_i^* = \int d{\bf a}\; 1 =1$, with analogous results for the $b$ integrals. 

Using these, we find that 
\begin{align}
{\cal{E}}(S)= & \frac{2}{d^2(d+1)^2}\left[\left|\sum_{ij}M_{ij,ij}\right|^2+d^2 \sum_{ijkl} \left|M_{ij,kl}\right|^2-d\sum_{ijkl}M_{ij,il}M^*_{kj,kl}
-d\sum_{ijkl}M_{ij,kj}M^*_{il,kl}\right] \nonumber \\
&  + O(M^3), \label{eq:epower_raw}
\end{align}
where in this expression, we have {\bf not} used Einstein summation convention.  The term in square brackets can be massaged into a form which is a sum of positive terms.  For the entanglement power to vanish, each of these terms must vanish separately, which impose the entanglement conditions Eqs.~\ref{eq:c1}-\ref{eq:c3}.  Specifically, we find
\begin{align}
{\cal{E}}(S)=  \frac{2}{d^2(d+1)^2}\Big[& d^2 \sum_{ijkl, i \neq k, j\neq l} |M_{ij,kl}|^2 +\label{eq:epower_1} \\ & \frac{d}{2} \sum_{ijkl, j \neq l} |M_{ij,il}-M_{kj,kl}|^2+
  \frac{d}{2} \sum_{ijkl, i \neq k} |M_{ij,kj}-M_{il,kl}|^2
+\label{eq:epower_2} \\ & \frac{1}{4}\sum_{ijkl} |M_{ij,ij}+M_{kl,kl}-M_{il,il}-M_{kj,kj}|^2 \Big]  + O(M^3). \label{eq:epower_3}
\end{align}
This can be proven by expanding the terms in the sum and comparing to Eq.~\ref{eq:epower_raw}.  From this form, we can see that the three types of constraints are imposed by the three lines in Eq.~\ref{eq:epower_1}, \ref{eq:epower_2}, and \ref{eq:epower_3}.  This expression is also useful if one were to look for parameter points where $\cal E$ was suppressed but not zero, where the  coefficients would determine the  relative importance of the three types of constraints.

\bibliographystyle{utphys}
\bibliography{references}

\end{document}